\def\year{2016}\relax
\documentclass[letterpaper]{article}
\usepackage{aaai16}
\usepackage{times}
\usepackage{helvet}
\usepackage{courier}
\usepackage{url}
\usepackage{graphicx}

\begin{document}
%
\title{Consensus of Dependent Opinions}
\author{Sujoy Chatterjee$^1$, Anirban Mukhopadhyay$^1$ and Malay Bhattacharyya$^2$\\
$^1$Department of CSE, University of Kalyani, Nadia -- 741235, India\\
E-mail: \{sujoy, anirban\}@klyuniv.ac.in\\
$^2$Department of IT, IIEST, Shibpur, Howrah -- 711103, India\\
E-mail: malaybhattacharyya@it.iiests.ac.in\\
}

\maketitle

\begin{abstract}
Providing opinions through labeling of images, tweets, etc. have drawn immense interest in crowdsourcing markets. This invokes a major challenge of aggregating multiple opinions received from different crowd workers for deriving the final judgment. Generally, opinion aggregation models deal with independent opinions, which are given unanimously and are not visible to all. However, in many real-life cases, it is required to make the opinions public as soon as they are received. This makes the opinions dependent and might incorporate some bias. In this paper, we address a novel problem, hereafter denoted as dependent judgment analysis, and discuss the requirements for developing an appropriate model to deal with this problem. The challenge remains to be improving the consensus by revealing true opinions.
\end{abstract}

\section{Introduction}
Opining about a large set of questions (may be images, tweets, etc.) within a bounded time is a challenging job. It highly demands for a distributed processing. Recently, it has been understood that such problems can be easily solved with human intervention and crowdsourcing is one such successful model \cite{Bhattacharyya2013} \cite{Chatterjee2015}, \cite{Kajino2012}, \cite{Lease2012}. Still, there remain a few challenges like assigning the appropriate domain experts as crowd workers, assuring the quality of opinions, etc. Due to the lack expertise or dishonest interest of crowd workers, the opinions have a high rate of noise \cite{Raykar2011}. To obtain a better consensus of the given opinions, aggregation models are frequently used \cite{Sheshadri2013}. In general, opinion aggregation models deal with independent opinions. These are given unanimously and are not visible to everyone. However, in many real-life scenarios, it is required to make the opinions public. The crowd workers should be able to choose the questions (in which they feel they have sufficient knowledge) to be answered. These are possible with an open crowdsourcing platform that makes the collection of opinions faster and transparent \cite{Bhattacharyya2014}. In contrast to investigating the impact of review bias as was studied earlier \cite{Krishnan2014}, we aim here to derive judgments from dependent opinions. The limited existing approaches in this direction based on incentive mechanisms for unknown objective truth \cite{Prelec2004} and debiasing methods for consensus tasks \cite{Kamar2015} are not appropriate for this open model problem.

If the opinions are made open to all the crowd workers, there is a high chance of inclusion of bias in the opinions. As the opinions become dependent, aggregating them introduces a new type of judgment analysis problem. Finding the consensus of dependent crowd opinions leads to a relatively new research direction of judgment analysis and there is hardly any study available in the literature. In this paper, we address this new research direction of dependent crowd judgment analysis problem. Principally, in this problem the opinions arrive with a time interval and become visible to all the crowd workers. So, the opinions provided are an ordered (dependent) set of opinions. As the crowd workers may notice others' opinions, therefore, some preconceived notions might influence their opinions.

If a crowd worker realizes that his true opinion (the opinion in mind) is wide dissimilar from the majority, he may change his opinion (the opinion disclosed) towards the majority. A crowd worker sticking to his actual opinion, even after observing others' opinions, establishes that he is confident about his opinion. On the contrary, if a crowd worker deviates from his true opinion by a high scale, it reflects that he is not much confident. A crowd worker who is less confident about his opinions might not be efficient. Therefore, how far a crowd worker differs from others should be taken into account. From this perspective, we discuss about some novel terms. These include drop of confidence (absolute difference of a crowd worker's actual and disclosed opinion), reliability, and accuracy (how dissimilar the crowd worker's opinion is from the rest of the workers) of the crowd worker's opinion. Ultimately, these can be used as weights to derive the aggregated judgment of a question.

\section{Motivation}
The problem of finding consensus of dependent opinions is basically motivated from a recently proposed crowd-powered reviewing model. This model collects the review score for research papers (submitted in video format) from crowd workers. Thus, it yields the opinions (annotation scores) of crowd reviewers on the said research materials. Using an online platform, the reviews are basically collected \cite{Bhattacharyya2014}. For this specific platform, the possible options for opinions are `strong accept', `accept', `weak accept', `borderline', `weak reject', `reject', and `strong reject'. However, such models can be used for any open crowdsourcing platform where the crowd workers provide a dependent nature of feedback.

\section{Problem Formulation}
\begin{figure}
\begin{center}
\includegraphics[height=1.4in]{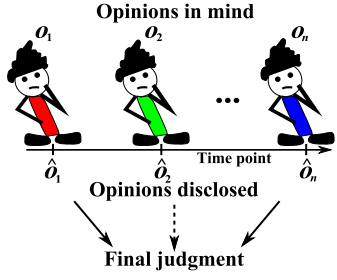}
\end{center}
\caption{Collecting dependent opinions from crowd workers. True opinion of the $i^{th}$ worker might be $O_i$ but it is disclosed as $\hat{O}_i$ (due to bias) after seeing the others' opinions $\{\hat{O}_1, \hat{O}_2, \ldots, \hat{O}_{i-1}\}$.}
\label{Fig:Dependent}
\end{figure}

Let us now formalize the dependent judgment analysis problem for a crowd-powered environment. Suppose we have a set of $m$ questions $Q = \{q_1, q_2, \ldots, q_m\}$ that are labeled by $n$ number of crowd workers $W = \{w_1, w_2, \ldots, w_n\}$. Further assume that the set of optional labels to be given as opinions is given by $L = \{l_1, l_2, \ldots, l_k\}$. Given these precursory assumptions, an annotation process is a quadruplet $(Q, W, \tau, T)$ consisting of a mapping function $\tau: Q \times W \rightarrow L \times T$ and a set of arrival times. The objective is to obtain the final judgment of all the questions in $Q$. Fig.~\ref{Fig:Dependent} depicts one such scenario where the dependent opinions arrive successively over time. Note that, $\hat{O}_1 = O_1$. Note that, the order of arrival of crowd workers is the only important factor here, not the distribution of arrival.

\section{Challenges}
It can be probabilistically understood that revealing the true opinion of a crowd worker from the given set of dependent opinions is not an easy task. Following the conditional probability, we have $P(O_i = \hat{o}_i|O_1 = \hat{o}_1, \ldots, O_{i-1} = \hat{o}_{i-1})$
\begin{equation}
= \frac{P(O_1 = \hat{o}_1, \ldots, O_{i-1} = \hat{o}_{i-1}|O_i = \hat{o}_i)}{P(O_1 = \hat{o}_1, \ldots, O_{i-1} = \hat{o}_{i-1})}.
\end{equation}

So, it might be useful to take the opinions in two different phases from the crowd workers to understand the dependence of the opinions from two different orders. Initially, the crowd workers provide their opinions independently (prior independent score) and these independent opinions are not disclosed to the others. In the next phase, all these independent opinions are made public so that every crowd worker can observe others' independent scores. Then again, the opinions are obtained (posterior dependent score) from them on the same questions. This might provide a better understanding about the distribution of the actual opinions.

We now introduce a few terms that might be useful for a better implementation of the said ideas in the form of a judgment analysis model.

\textbf{Drop of Confidence:} The \emph{drop of confidence} of a crowd worker is the absolute difference between the prior independent score and posterior dependent score. If the mean score changes (i.e., majority of the crowd workers deviate from their initial opinions), any arbitrary crowd worker is likely to change his posterior dependent score. So, how a crowd worker is modifying his own score with respect to the others for a particular question can be expressed as the ratio of deviation of individual score with respect to the deviation of mean score in independent and dependent situations, respectively. The reliability of a crowd worker can be quantified as the reciprocal of this ratio.

\textbf{Reliability:} The \emph{reliability} of a crowd worker can be defined in terms of the drop of confidence of a crowd worker in comparison with the mean deviation of two different situations (i.e., independent and dependent). It reflects how much confident the crowd worker is over his own opinions. When a crowd worker changes his score drastically after viewing others' scores, he might not be confident enough. Thus, some bias gets included into his dependent opinion. On the other side, if a crowd worker sticks to his own score even after seeing the independent scores of others, the crowd worker is confident about his opinion. However, only becoming confident, while giving an opinion, does not guarantee that the worker is good. The absolute difference between his posterior dependent score and the mean value of all the posterior dependent scores should also be less. This gap highlights the closeness of a crowd worker with others on the basis of their posterior dependent scores.

\textbf{Accuracy:} The \emph{accuracy} of a crowd worker is the closeness of his posterior dependent score to the mean of all posterior dependent scores over a particular question. This means how much similar a crowd worker's opinion is with rest of the crowd workers. A better accuracy score indicates that the crowd worker is highly similar to the other crowd workers. So, increasing this accuracy score with a less drop of confidence means a crowd worker is reliable and accurate.

\section{Conclusions}
We introduce a new judgment analysis problem that considers both independent and dependent opinions of crowd workers. We also discuss the requirements for developing an appropriate consensus model for managing dependent opinions of crowd workers along with independent crowd opinions. Consideration of the ambiguity of questions (for which confidence may drop) and the self-reported confidence scores of crowd workers (collected with annotations) can make the model more robust. It is also interesting to consider that a subset of existing opinions are only available, instead of all the opinions. Consideration of a paid crowdsourcing model is also encouraging.

\section*{Acknowledgments}
The work of Malay Bhattacharyya is supported by the Visvesvaraya Young Faculty Research Fellowship 2015-16 of DeitY, Government of India. All the authors would like to thank the crowd contributors involved in this work.

\bibliographystyle{aaai}
\bibliography{Reference}

\end{document}